\documentclass{WileyMSP-template}
\usepackage[utf8]{inputenc}
\usepackage[english]{babel}
\usepackage[T1]{fontenc}
\usepackage{fancyhdr}
\usepackage{graphicx}
\usepackage{abstract}
\usepackage{authblk}
\usepackage{setspace}
\usepackage{geometry}
\usepackage{float}
\usepackage{upgreek}
\usepackage{caption}
\usepackage{subcaption}
\begin{document} 
	%%%%%%%%%%%%%%%%%%%%%%%%%%% TITLE %%%%%%%%%%%%%%%%%%%%%%%%%%%%%%%%%%%%%%%%%%%%%%%%%%%%%%%
	\title{Wafer-Scale Fabrication of Hierarchically Porous Silicon and Silica Glass by Active Nanoparticle-Assisted Chemical Etching and Pseudomorphic Thermal Oxidation\\}
	%Chemical etching guided by catalitically propelled nanoparticles and subsequent thermal oxidation results in wafer-scale hierarchical porosity in single-crystalline silicon and silica glass
	
	\maketitle
	%\newline
	%%%%%%%%%%%%%%%%%%%% Authors and Affiliation %%%%%%%%%%%%%%%%%%%%%%%%%%%%%%%%%%%%%%%%%%
	% Affiliations: Please provide adacemic titles (Prof. or Dr.) for all authors where applicable, and include an institutional email address for all corresponding authors
	%
	\author{Stella Gries*}
	\author{Manuel Brinker}
	\author{Berit Zeller-Plumhoff}
	\author{Dagmar Rings}
	\author{Tobias Krekeler}
	\author{Imke Greving}
	\author{Patrick Huber*}
	%
	%\newline
	%
	
	\begin{affiliations}
		\textbf{Stella Gries}\\
		Institute for Materials and X-Ray Physics, Hamburg University of Technology, Denickestr. 10, 21073 Hamburg, Germany;\\ 
		Center for X-Ray and Nano Science CXNS, Deutsches Elektronen-Synchrotron DESY, Notkestr. 85, 22607 Hamburg, Germany, and\\ 
		Centre for Hybrid Nanostructures CHyN, University of Hamburg, 22607 Hamburg, Germany.\\
		*Email Address: stella.gries@tuhh.de\\
		ORCID-ID: 0000-0002-0869-1665\\
		\vspace{0.5cm}
		\textbf{Manuel Brinker}\\
		Institute for Materials and X-Ray Physics, Hamburg University of Technology, Denickestr. 10, 21073 Hamburg, Germany;\\ 
		Center for X-Ray and Nano Science CXNS, Deutsches Elektronen-Synchrotron DESY, Notkestr. 85, 22607 Hamburg, Germany, and\\ 
		Centre for Hybrid Nanostructures CHyN, University of Hamburg, 22607 Hamburg, Germany.\\
		\vspace{0.5cm}
		\textbf{Dr. Berit Zeller-Plumhoff}\\
		Center for X-Ray and Nano Science CXNS, Deutsches Elektronen-Synchrotron DESY, Notkestr. 85, 22607 Hamburg, Germany, and\\
		Institute of Metallic Biomaterials, Helmholtz Zentrum Hereon, 21502 Geesthacht, Germany.\\
		\vspace{0.5cm}
		\textbf{Dagmar Rings}\\
		Electron Microscopy Unit, Hamburg University of Technology, 21073 Hamburg, Germany.\\
		\vspace{0.5cm}
		\textbf{Dr. Tobias Krekeler}\\
		Electron Microscopy Unit, Hamburg University of Technology, 21073 Hamburg, Germany.\\
		\vspace{0.5cm}
		\textbf{Dr. Imke Greving}\\
		Center for X-Ray and Nano Science CXNS, Deutsches Elektronen-Synchrotron DESY, Notkestr. 85, 22607 Hamburg, Germany, and\\
		Institute of Materials Physics, Helmholtz Zentrum Hereon, 21502 Geesthacht, Germany.\\
		\vspace{0.5cm}
		\textbf{Prof. Dr. Patrick Huber}\\
		Institute for Materials and X-Ray Physics, Hamburg University of Technology, Denickestr. 10, 21073 Hamburg, Germany;\\ 
		Center for X-Ray and Nano Science CXNS, Deutsches Elektronen-Synchrotron DESY, Notkestr. 85, 22607 Hamburg, Germany, and\\ 
		Centre for Hybrid Nanostructures CHyN, University of Hamburg, 22607 Hamburg, Germany.\\
		*Email Address: patrick.huber@tuhh.de\\
	\end{affiliations}
	%
	% Keywords: Please provide a minimum of three and a maximum of seven keywords, separated by commas
	%

	%%%%%%%%%%%%%%%%%%%%%%%%  START %%%%%%%%%%%%%%%%%%%%%%%%%%%%%%

	%\date{\begin{flushleft}\today\end{flushleft}}
	
	%\vspace{-4cm}
	
	%\maketitle 
	%\vspace{-0.5cm}
	\newpage
	%%%%%%%%%%%%%%%%%%%%%% ABSTRACT %%%%%%%%%%%%%%%%%%%%%%%%%%%%%%%%%
	\begin{abstract}
		Many biological materials exhibit a multiscale porosity with small, mostly nanoscale pores as well as large, macroscopic capillaries to simultaneously achieve optimized mass transport capabilities and lightweight structures with large inner surfaces. Realizing such a hierarchical porosity in artificial materials necessitates often sophisticated and expensive top-down processing that limits scalability. Here we present an approach that combines self-organized porosity based on metal-assisted chemical etching (MACE) with photolithographically induced macroporosity for the synthesis of single-crystalline silicon with a bimodal pore-size distribution, i.e., hexagonally arranged cylindrical macropores with 1 micrometer diameter separated by walls that are traversed by mesopores 60 nm across. The MACE process is mainly guided by a metal-catalyzed reduction-oxidation reaction, where silver nanoparticles (AgNPs) serve as the catalyst. In this process, the AgNPs act as self-propelled particles that are constantly removing silicon along their trajectories. High-resolution X-ray imaging and electron tomography reveal a resulting large open porosity and inner surface for potential applications in high-performance energy storage, harvesting and conversion or for on-chip sensorics and actuorics. Finally, the hierarchically porous silicon membranes can be transformed structure-conserving by thermal oxidation into hierarchically porous amorphous silica, a material that could be of particular interest for opto-fluidic and (bio-)photonic applications due to its multiscale artificial vascularization. 
	\end{abstract}
	
	%%%%%%%%% INTRODUCTION %%%%%%%%%%%%%% 
	
	\section{Introduction}
	Hierarchical porosity is a key architectural principle in many biological tissues and biomaterials to achieve simultaneously leightweight structures, large inner surfaces and homogeneous vascularization in combination with optimized mechanical robustness and mass transport capabilities \cite{Murray1926,Zheng2017,Yang2017HierarchicalPorousMedia, Eder2018}. Realizing such a hierarchical porosity in artificial porous materials is therefore a very active research field, in particular with regard to scalable fabrication methods \cite{Yang2017HierarchicalPorousMedia, Wang2018, Jung2022,Huo2020, Condi2022, Putz2018, Putz2020, Rebber2022, Sai2013}. Inspired by recent advances in the synthesis of multiscale porous metals\cite{Shi2021} and ceramics\cite{Huo2020, Putz2018, Hwang2018, Bilo2020}, the main goal of this study is a scalable fabrication of the mainstream semiconductor silicon with open macro- and nanoporosity.
	
	Electrochemically induced nanoporosity in silicon (np-Si) has been intensively studied both with regard to fundamental and applied aspects after its accidental discovery at Bell labs in 1956 \cite{Sailor2009, Canham2018, Uhlir1956}. It results in specific surface areas of 200 - 500 m$^{2}$/g \cite{Lehmann1991,Herino_1987,Harraz2014,Lehmann2005}, tailorable by the fabrication parameters \cite{Kumar2007a, Kumar2008a}. Nanoporous Si as a nanostructured scaffold offers a wide range of applications for the exploration of the fundamentals of spatially confined condensed-matter systems \cite{Vincent2014, Henschel2007, Huber2015,Gor2015,Gor2017, Bossert2021}. As a nanoporous mainstream semiconductor np-Si is interesting for sensorics, actuorics, (bio-)photonics and drug delivery \cite{Sailor2009,Geng2011,Brinker2022, Cencha2020, Makila2019, Alhmoud2021, Gang2020, Balzer2019, Martinez2013} as well as energy storage and conversion \cite{Tian2007,Li2015,Dai2019,Bang2011, Chen2018}. Filling the pore space with soft or hard materials, for example, electrically conductive polymers can improve properties such as electrical conductivity and mechanical robustness \cite{Huber2015,Bang2011,Harraz2011,Westover2014,Thelen2021, Brinker2020, Gostkowska2022}.\\ 
	
	For the fabrication and in-operando performance of many nanoporous, functional (hybrid) materials based on np-Si an effective and fast mass transport in pore space is important. However, in the extreme spatial confinement of np-Si, the pure spatial restriction and the interactions with the pore walls can hamper significantly hydrodynamic flow and self-diffusion, which results in slow process kinetics \cite{Brinker2022,Brinker2022HowSimulations, Brinker2020,Schlaich2017,Gupta2020,Gonella2021,Kusmin2010,Kusmin2010a}. \\
	
	One approach to address this challenge is the synthesis of hierarchically porous silicon (hp-Si). Pore diameters in the order of micrometers result in bulk-like flow properties and a low diffusion inhibition. Thus, a material of coexisting pores of different sizes is an approach to significantly improve the transport performance and ease the application \cite{Berdys2015}.\\
	
	Recently developed hierarchical structures in silicon, such as presented in \cite{Makila2019,Alhmoud2021,Bang2011,Berdys2015} and \cite{Lehmann2005} consist mainly of porous nanoparticles or dead-end pores. Here, we follow the strategy to start with a photolithographically fabricated macroporous, monocrystalline, free-standing silicon membrane and employ then a self-organized nano-porosification of the macroporous scaffold structure. Thus, we combine a top-down method with self-propelling particle processing. Up to now, for the formation of nanopores in silicon mainly anodic dissolution in an electrolyte containing hydrofluoric acid is utilised\cite{Lehmann1991}. In this process, the etching direction and rate for monocrystalline silicon always depends on the crystal orientation and the doping concentration. The pores thus produced exhibit an anisotropic collective orientation.\\
	We resort here to an alternative etching process, which is supposed to ensure the dissolution of silicon without application of external electrical potentials and which is independent of the crystal orientation \cite{Huang2011,Geyer2013,Sailor2011,Han2014,Pinna2020,Peng2008}. The electroless etching process is catalyzed by metallic particles, in this case silver nanoparticles (AgNPs). Metal-assisted chemical etching (MACE) has been mainly used so far for the synthesis of silicon nanowires (SiNW) \cite{Huang2011,Peng2006,Wendisch2020,Chen2017a, Chen2017} and only a few studies address the synthesis of porous silicon structures \cite{Li2000}.\\ 
	
	As we will demonstrate in the following the resulting hp-Si membranes can be large with an area limited only by the size of the initial wafer and a thickness of $\mathrm{200-300\,\mu m}$. Moreover, they can be  pseudomorphic transformed with barely any changes in the multiscale porosity into monolithic, amorphous hierarchically porous silica (hp-SiO$_2$) via thermal oxidation. 
	
	%%%%%%%%%%%%%%%% RESULTS AND DISCUSSION %%%%%%%%%%%%%%%%%
	
	\section{Results and Discussion} 
	
	\subsection{Evolution of hierarchical porosity in macroporous silicon by catalytic, silver nanoparticle-guided etching}
	\begin{figure}[H]
		\RaggedRight
		\includegraphics[width=1\textwidth]{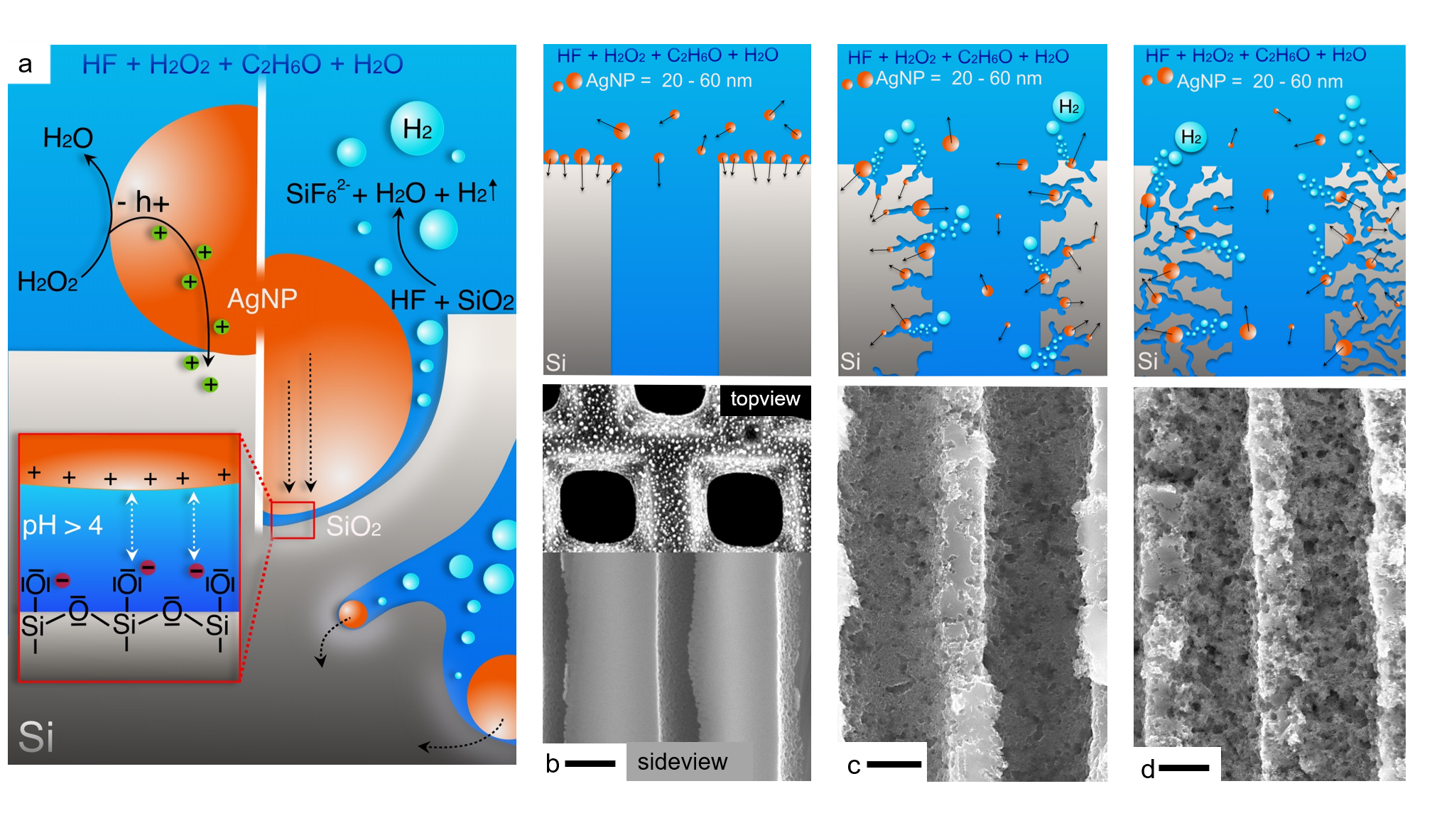}
		\caption{Silver nanoparticle-assisted chemical etching of porous silicon as a function of increasing etching time. Orange spheres indicate the silver nanoparticles (AgNPs), the light blue background is the etching solution in which the membranes are immersed. Grey areas mark the silicon pore walls. (a) Schematic illustration of a single AgNP etching process at a silicon surface, where on the left the reduction-oxidation reaction is taking place and on the right the following dissolution of silica in HF containing solution. (b) upper picture: Side-view on one channel of the initial macroporous Si membrane with AgNPs deposited on the outer surface, lower pictures: electron micrograph of the initial macroporous membrane in side-view and precipitated AgNP at the pore openings in top-view, (c) hp-Si structure after 20 min etching time: The AgNPs are migrating into the macropore space, attach to the Si-surface and induce locally the formation of mesoscaled dead-end pores, (d) hp-Si after 45 min etching time: The AgNP particles perform long walks through the pore walls and the formed channels merge and break through which interconnects the macropores. (scale bar $=$ 500 nm)}
		\label{fig:MACE}
	\end{figure}
	
	Monocrystalline macroporous silicon membranes doped with phosphorus (n-type) serve as the starting material. The membranes have a <100> crystal orientation parallel to the surface, a pore diameter of 1 $\upmu$m and a pore pitch, i.e. distance from pore centre to pore centre, of 1.5 $\upmu$m. The macropores are arranged hexagonally in the membrane with their long axes parallel to the membrane surface. The samples are 10 x 10 mm$^2$ in size and 0.2 - 0.22 mm in thickness.
	
	For the MACE-process the membrane is first coated with AgNPs. The deposition of the silver particles is performed via precipitation from 2 mM aqueous silvernitrate solution for 20 minutes. The silver ions in the solution are reduced on the silicon surface to elementary silver by taking up an electron from silicon at the surface. This electron can be provided from the phosphorous doping in the silicon or from the negatively charged silicon-dioxide layer at the surface which attracts the positively charged silver-ions.\\
	It is remarkable that independent from the macropore size used the silver is deposited on the outer surface just a few micrometers inside the pore openings of the membrane, but not in the central area of the macropores. The deposited AgNPs are annealed at 400°C in an oxygen free atmosphere for at least 2 hours to adjust the shape and size of the particles. \\
	After annealing the membrane is immersed in the etching solution which consists of a 1:1:1 mixture of hydrofluoric acid (HF 48\%), ethanol (C$_2$H$_6$O 99,99\%) and hydrogen peroxide (H$_2$O$_2$ 30\%). The resulting porosity can be adjusted by varying the etching time.\\
	
	In the lower section of Figure~\ref{fig:MACE} electron micrographs are shown which document the evolution of the pore structure in silicon during increasing MACE time while concentrations of HF, H$_{2}$O$_{2}$ and AgNP are kept constant. In the upper section we illustrate how we envision the porosification process as a result of the collective self-propelled movements of the AgNP particles and the subsequent dissolution of silicon material along their trajectories in the scaffold structure. The first image (b) is the initial silicon membrane with straight macroporous through-pores running from top to bottom of the image and solid pore walls. In the corresponding upper panel the deposited AgNP are indicated by the orange circles, which can not be seen in the side-view SEM-image, but in the top-view of Figure~\ref{fig:MACE}b. The electron micrographs in (c) and (d) show how etching with AgNP leads to higher porosity with increasing exposure time. Additionally, the pore  diameters of both the initial macropores and the synthesized mesopores increase with longer etching times, since the pores are widening in other words the pore walls are thinning out. The images also illustrate the increase of the inner surface area and thus the increasing roughness and porosity. 
	
	The etching results show that the AgNP serve as catalysts, they do not incorporate into or remain in the pore walls. They are constantly drilling into the silicon and dissolve the membrane selectively along their trajectories as long as the membrane is immersed in the etching solution. It is not fully understood if the AgNPs are flushed out from the going-through pores by H$_{2}$-gas evolution or if they are dissolved in the HF-solution to ions or intermdeiate molecules like AgF. \\
	
	At first glance, this behavior of the AgNPs in the MACE process is very reminiscent of self-propelled Brownian Janus particles or nanoswimmers having catalytic active and inactive sides for H$_{2}$O$_{2}$ \cite{Paxton2004,Bechinger2016,Shaebani2020, Gompper2020}. In a solution of H$_{2}$O$_{2}$, the catalytic oxidation-reduction reaction at one particle side induces fluid flow along the surface through self-diffusiophoresis. The resulting hydrodynamic drag on the AgNP results in a directed particle motion which dominates over the random thermal self-diffusion, as one can explore in relatively simple optical microscopy experiments \cite{Paxton2004}. Here these processes occur within the opaque Si scaffold under harsh electrochemical conditions, so that the dynamics can only be indirectly inferred from such electron micrographs recorded after certain processing times, as shown in Fig. \ref{fig:MACE}.
	
	Note however that the situation in the MACE process is much more complex. The propulsion process inside the silicon appears in a highly spatially confined geometry and the symmetry is not broken by an intrinsic asymmetry of the particle, but rather by the spatial confinement or (multiple) particle-silicon surface contact(s). Here the contact with the solid surfaces may result in an additional quenching of Brownian rotation or selection of distinct rotation modes and thus result in much more directed motion than in classical self-propelled colloidal particles in bulk aqueous solutions \cite{Das2015}. To make the situation even more complex, the excess number of electron holes (h$^+$) can migrate in silicon to locations far away from the injecting AgNP and result there in an additional silicon dissolution and hydrogen evolution. We illustrate these coupled processes in Figure~\ref{fig:MACE}d. There is a possibilty that holes are also recombining with electrons from the phosphorous doping, but since the amount of injected holes is a tremendous excess, there are always enough for etching. It would be a study for itself to investigate the doping concentration needed to tune the injected hole concentration within the material. Additionally, solution concentration gradients could contribute via diffusiophoresis to the directed motion of the AgNPs in the dead-end pore geometry \cite{Battat2019}. To the best of our knowledge these remarkably complex, multi-chemo-physical couplings have been barely explored so far. Also the material removal patterns, even for the much simpler planar silicon geometries are a matter of controversial debates \cite{Huang2011,Peng2008,Magagna2020,Miao2017,Lee2015,Tsujino2005}. 
	
	In fact, it appeared to us that one important aspect has been completely ignored so far in the MACE literature, namely a potential attraction of the AgNP to the silicon oxide that is forming at the AgNP/silicon interface. Since the pH value in the MACE process is larger than the isoelectric point of silica (pH$_{iso}$=3.9), the Si-OH groups of this interface are partially deprotonated. The surface is negatively charged. 
	There are two possible attractive forces which could guide the AgNP through the silicon. First, the negative charge could induce polarised layers in the electrolyte and lead to a positive mirror-charge in the AgNP which attracts the particle towards the solid surface. Second, we assume that the AgNP could be ionised, at least at the outer surface, by the HF-containing etching solution and ionic attractive forces lead to a movement of the AgNP towards the silicon. Metallic particles move into and through silicon, this can be observed from several experiments and articles, but how and why they are moving is not discussed so far.

	The interplay of this electrostatic pinning effect with the conventional hydrodynamic propulsion forces due to gas evolution and concentration gradients along with the self-confining path formation and the hydrodynamic interaction with the corrugation and roughness of the confining walls \cite{Chase2022} may then also explain the complex, sometimes even spiral-like trajectories and thus pore structures observable in electron micrographs recorded for samples formed by MACE.
	
	Still, despite a missing in-depth understanding of the self-organized pore-formation process our experiments demonstrate that the synthesis of hp-Si via silver catalyzed etching allows a control of the resulting porosity by systematic adjusting the fabrication parameters. The most important manufacturing parameters that can be changed for individual structure adjustment are the etching time of a membrane, the composition of the etching solution and the concentration of deposited silver particles, where the last two are not altered in the study presented here. 
	The etching time and thus also the etching rate with which the mesopores in the macropore walls are synthesized depends on the amount and size of deposited AgNP and the ratio of HFaq 48\% and H$_{2}$O$_{2}$aq 30\%. In the described procedure, the reactants were used in a ratio of 1:1:1. By applying MACE to a prestructured membrane an isotropic sponge-like structure is formed.
	In SEM-EDX measurements no residual AgNPs nor traces of silver can be found in the silicon matrix after etching. The silver does not remain in the matrix as it is reported for etching attempts on bulk silicon \cite{Tsujino2005,Li2016}, so no additional cleaning step is necessary to remove the silver.

	\subsection{Pseudomorphic transformation of hierarchically porous silicon to hierarchically porous silica by thermal oxidation}
	It has been known for a long time that electrochemically synthesized mesoporous silicon can be completely converted into amorphous silica glass via thermal oxidation, while most of the existing porosity is retained \cite{Pap2005, Kityk2008, Sentker2018}. Is this also true for the mesoporous walls in hp-Si presented here?
	
\begin{figure}[H]
		\centering
		\includegraphics[width=0.6\textwidth]{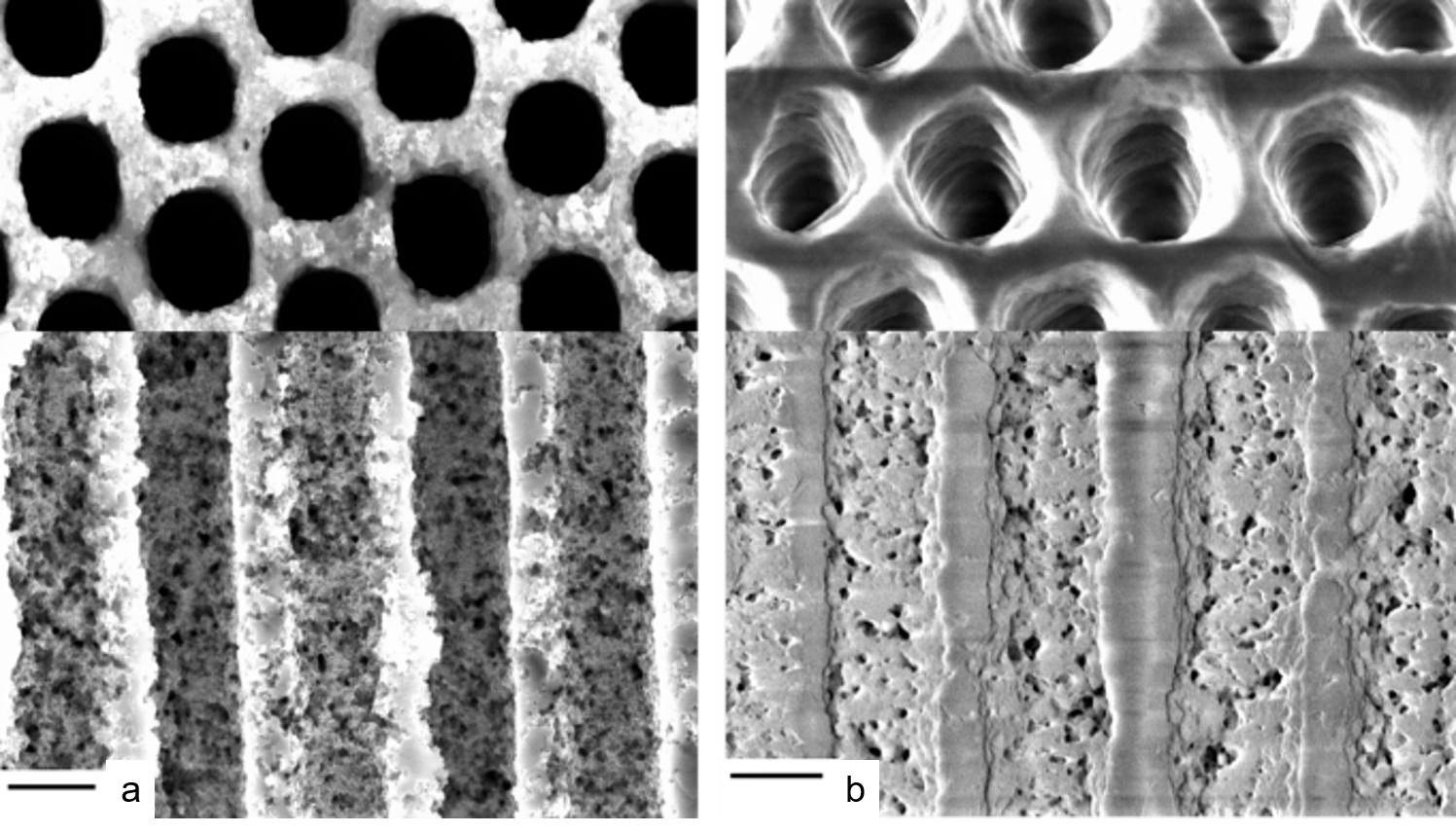}
		\caption{Comparison of hierarchically porous silicon and pseudomorphically transformed silica. Scanning electron micrographs of hp-Si (a) and hp-SiO$_{2}$ (b), surface and pore openings are shown in the upper row while the lower row displays the cross sections of the macropores. The scale bars indicate 1 $\upmu$m.}.
		\label{fig:therm_ox}
	\end{figure}

	For the transformation to silica, the hp-Si is stored at 800 - 1100 °C in a furnace in oxygen-containing ambient atmosphere for at least 48 hours, or longer depending on the thickness of the pore walls produced in the MACE steps.
	Figure~\ref{fig:therm_ox} shows SEM images of the top surface and cross-sections of the membranes, the first (left) is the hp-Si membrane and the second (right) is the resulting hp-SiO$_{2}$ structure. It can be clearly seen that the macropores, which extend from the top of the image to the bottom, have a reduced internal surface area and smaller pore diameters in the walls after oxidation. Wetting experiments show that the mesopores in the hp-SiO$_{2}$ still connect the macropores. This still allows lateral mass transport through the membrane. The pore size distribution also becomes narrower as smallest pores disappear and largest pore diameters shrink due to expansion of the pore walls. In the SEM images in Figure~\ref{fig:therm_ox}, it is also easy to see how the internal roughness of the macropore walls is reduced by oxidation.

\begin{figure}[H]
		\centering
		\includegraphics[width=0.8\textwidth]{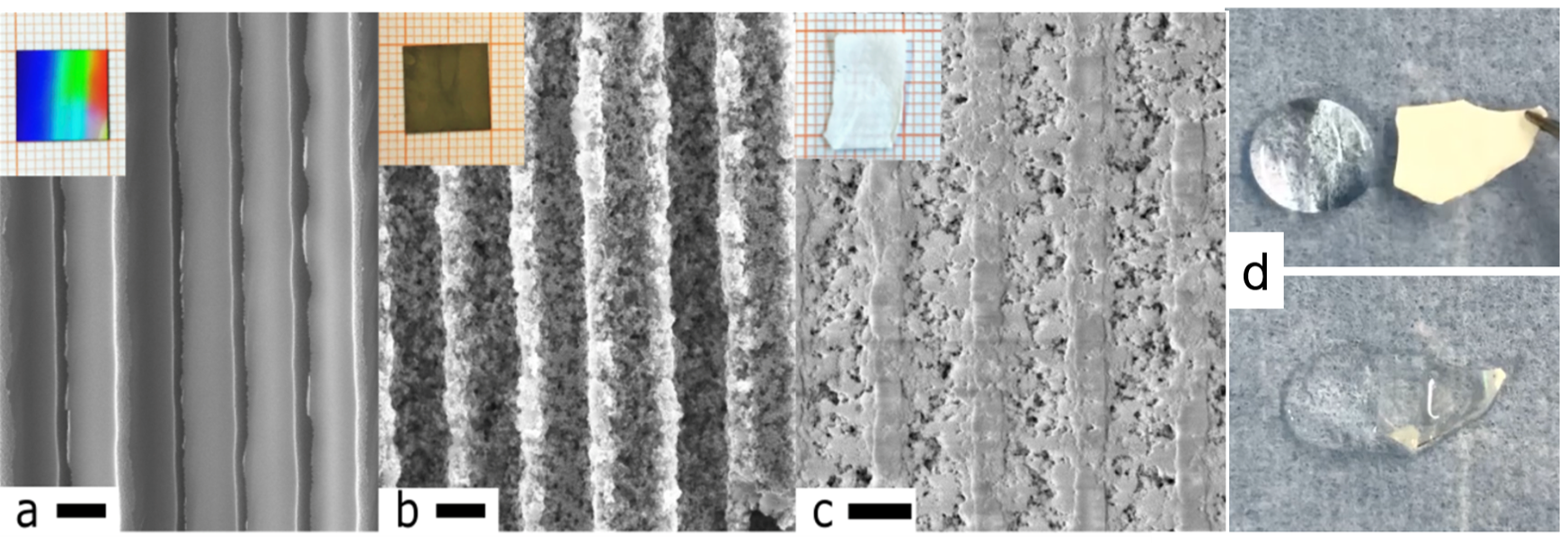}
		\caption{Transformation of a macroporous silicon membrane (a) via MACE to hierarchically porous silicon, hp-Si (b) and by pseudomorphic, thermal oxidation to hierarchically porous silica, hp-SiO$_2$ (c). Insets: Photographs of the 1x1 cm$^{2}$ membranes for each step (scale bar = 1 $\upmu$m). (d) Top-view on an as-prepared, dry and white hp-SiO$_2$ membrane piece next to a droplet of glycerin; (bottom) in comparison to the same matrix after bringing in contact with the droplet and subsequent spontaneous liquid imbibition. See also the Supplementary for a movie on the capillarity-driven spreading of the silica's refractive index-matching liquid and the corresponding dynamic changes in the optical transparency of hp-SiO$_2$.}
		\label{fig:transform}
	\end{figure}

	The two transformation steps from macroporous Si to hp-Si and to hp-SiO$_{2}$ are shown in Figure~\ref{fig:transform}. The membranes change their visual appearance with each treatment and the SEM-image shows how the pore morphologies change. 
	
	During oxidation, oxygen diffuses into the structure, thus amorphous silica is formed from monocrystalline silicon. This embedding of oxygen results in a macroscopic volume increase of up to 10\%. This expansion is less than previously observed expansions of dense, bulk silicon during oxidation, which can be more than 50\% \cite{Avouris1997}.
	One reason for the lower expansion is the special hierarchically porous structure. The oxidation process leads to a densification of the material, i.e. the porosity decreases, the pores become smaller and the smallest pores disappear as the opposing pore walls fuse together. Since the material offers a lot of free volume for densification during oxidation due to its large inner surface area and high porosity, the expansion of the entire membrane is less than in bulk material.\\
	Note that the dry hp-SiO$_2$ membrane appears white due to light scattering at pore wall/air interfaces with distances comparable to the wavelength of visible light. By filling the pores with a liquid that matches the refractive index of the silica walls this scattering can be completely suppressed. This is demonstrated by capillarity-driven, spontaneous imbibition experiments with liquid glycerin that matches the refractive index of silica ($n=$1.4). Evidencing the complete open porosity of the synthesized material the light scattering in the liquid-infused material is then completely suppressed and the matrix appears as optically transparent as fused silica, see Fig. \ref{fig:transform}d and the movie in the Supplementary.  
	
	\subsection{Quantitative multiscale porosity analysis}
\begin{figure}[H]
		\centering
		\includegraphics[width=0.8\textwidth]{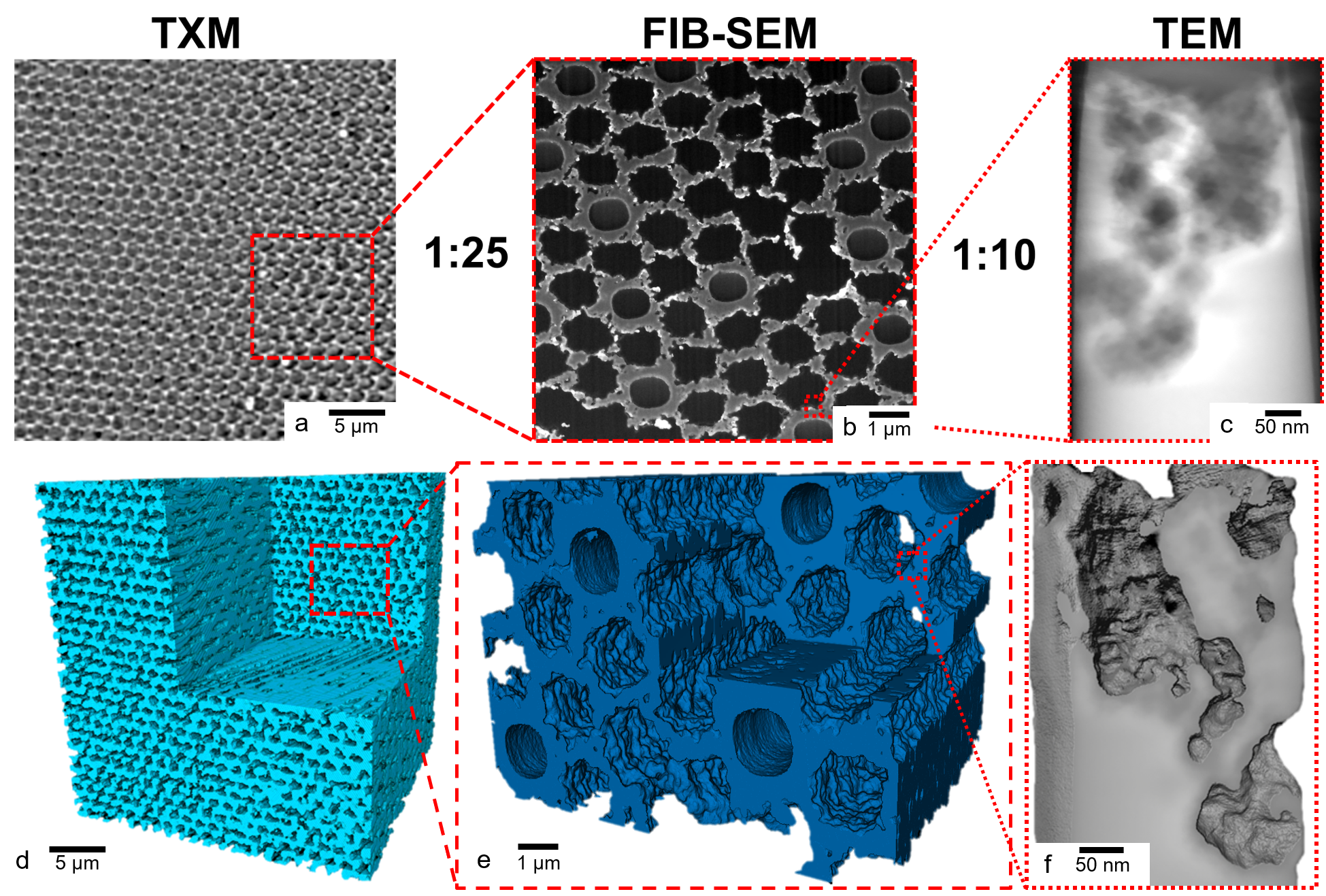}
		\caption{2D- and 3D-Visualization of the multiscale hp-Si pore space. (a-c) 2D images and (d-f) 3D sample reconstructions of the hp-Si structure on different length scales from (a) biggest to (c) smallest. (a,d) Macropores reconstructed from synchrotron-based transmission X-ray microscopy (TXM), (b,e) Focused ion beam scanning electron microscopy (FIB-SEM) representation of interconnection of the macropores, (c,f) transmission electron microscopy (TEM)-based tomography of a single mesopore in a macropore wall. In the Supplementary corresponding tomography movies are available.}
		\label{fig:2D3D}
	\end{figure}

	Using transmission X-ray microscopy (TXM), focused-ion beam scanning electron microscopy (FIB-SEM) and transmission electron tomography (TEM tomo), the porosity of hp-Si was assessed across multiple scales.\\ 
	It was attempted to double-check the results for the structural characterization with classical physisorption experiments like BET or BJH, but the bigger structural features are out of the measurement range which leads to misleading and incomplete datasets. The combination of imaging experiments allows to draw a full picture of the hierarchical structure. \\
	The TXM revealed that the macropores are aligned in a triangular lattice with a pore distance (centre to centre) of 1.60 $\pm$ 0.14 $\upmu$m (mean $\pm$ std dev). The hierarchical structure is visualized in Figure~\ref{fig:2D3D} in the methods section from macropores down to a single mesopore. From the experimental 2D imaging results 3D reconstructions are made to visualize the interconnections of the hierarchically porous materials.
	
\begin{figure}[H]
		\centering
		\begin{subfigure}[b]{0.48\textwidth}
			\centering
			\includegraphics[width=\textwidth]{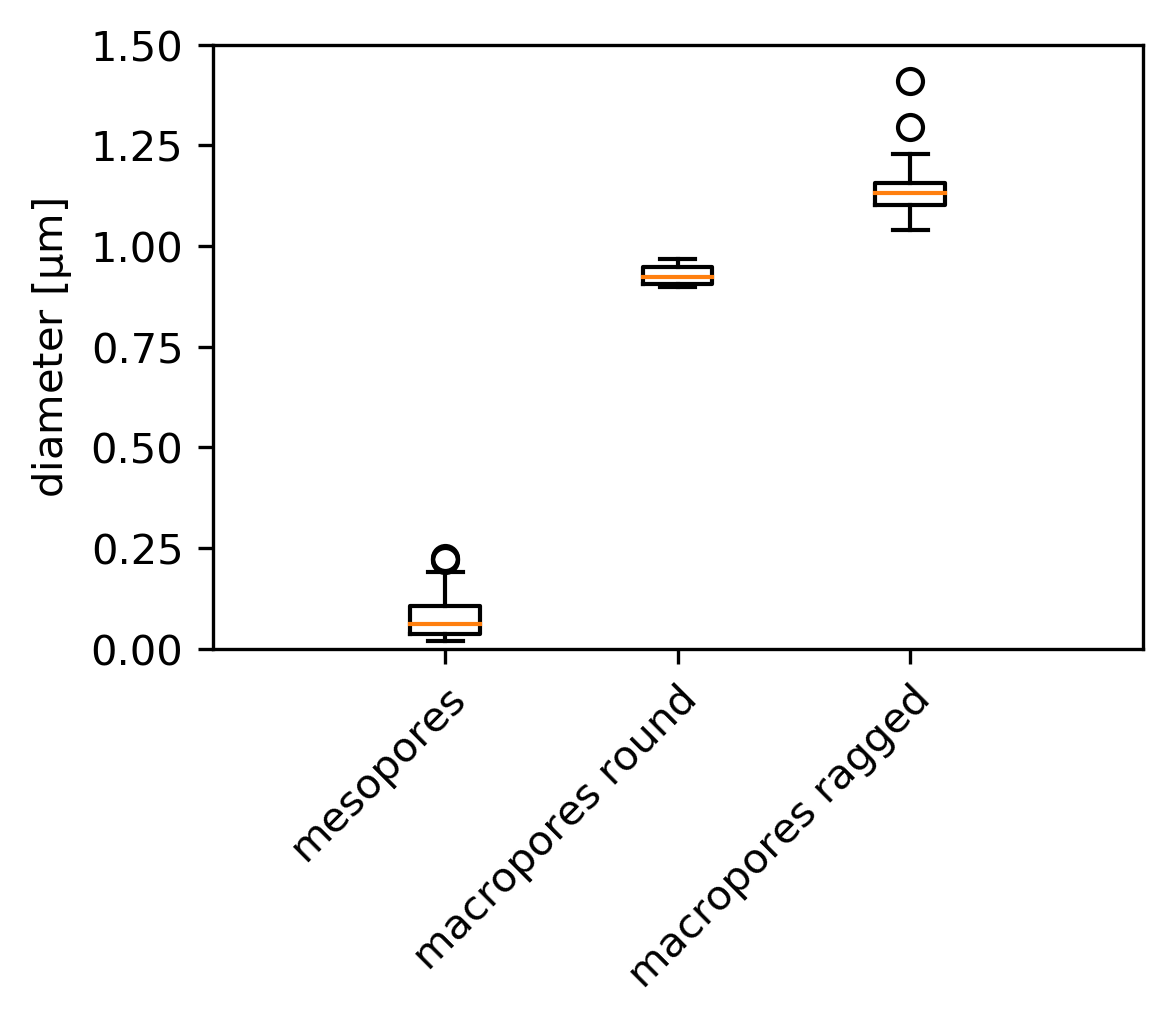}
			\caption{Boxplot visualization of the diameter distribution for all three pore types.}
		\end{subfigure}
		\hfill
		\begin{subfigure}[b]{0.48\textwidth}
			\centering
			\includegraphics[width=\textwidth]{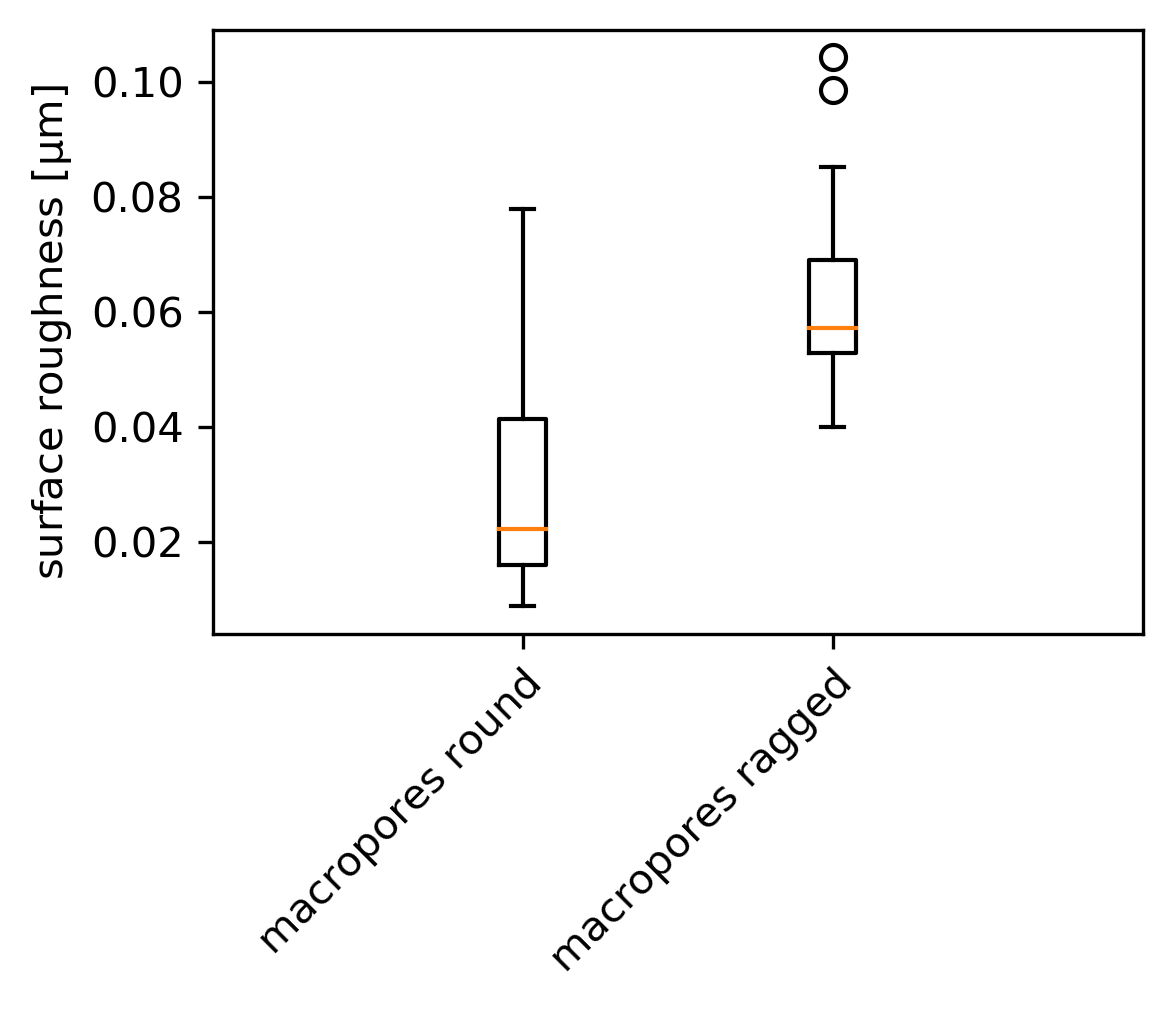}
			\caption{Boxplot visualization of the surface roughness Ra distribution for the macropores.}
		\end{subfigure}
		\caption{Quantified morphological parameters describing the macro- and mesopores based on FIB-SEM images.}
		\label{fig:morph_params}
	\end{figure}

\begin{figure}[H]
		\centering
		\includegraphics[width=0.9\textwidth]{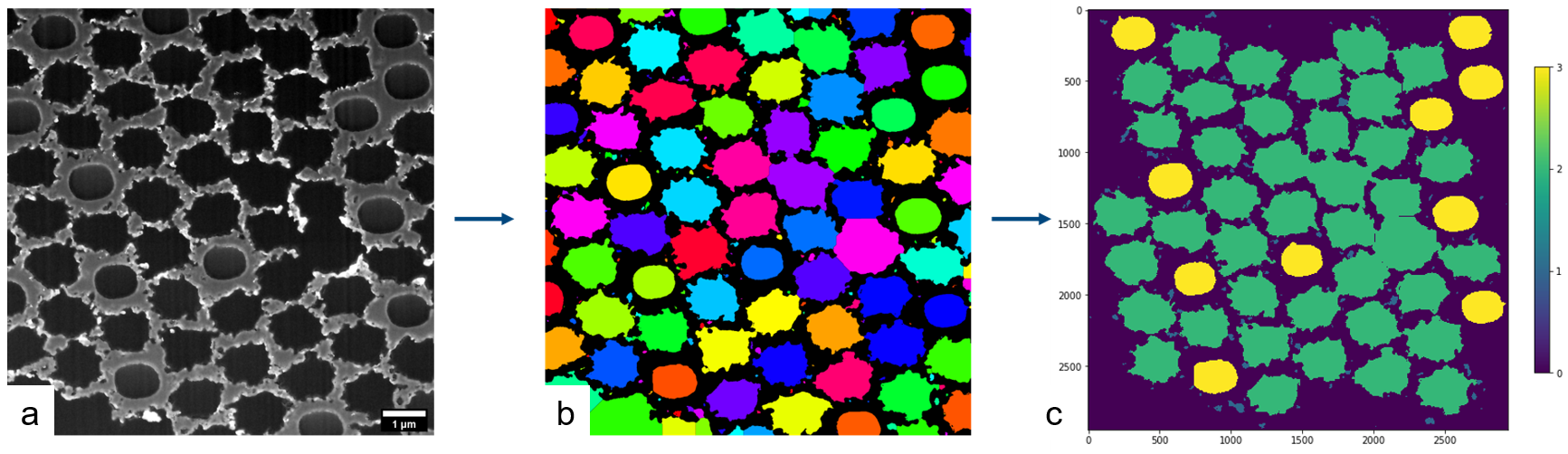}
		\caption{(a) FIB-SEM image following filtering using an iterative non-local means filter. The scale bar is 1 $\mu$m. (b) Labelled segmented pores within FIB-SEM image. (c) Pores classified according to their morphology. Pores in contact with the image. Borders were excluded from the analysis}
		\label{fig:workflow}
	\end{figure}

	The pore structure is revealed in more detail in FIB-SEM images, which allows for a differentiation of macropores into "round" and "ragged" macropores based on their appearance (see Figure~\ref{fig:workflow}). Some pores seem to be untouched by the etching (round) where other structures appear to be collapsed (ragged). Those 2D images show just the surface of the material and a few microns in depth. With regard to the full thickness of a membrane with 200 $\upmu$m the collapsed structures are just observed at the surface where strong gas-evolution can lead to ripped off areas whereas in the middle of the membrane is intact. The FIB-SEM further enables the visualization of the mesopores. To assess the macro- and mesoscale porosity, the FIB-SEM image is segmented and the pores are labeled individually. Subsequently, they are classified automatically based on their area and surface length using k-means clustering (k=3). This process and the classification outcome is visualized in Figure~\ref{fig:workflow}. Based on the classification, we have obtained the main parameters describing the pore morphology. Specifically, we calculated the pore diameter and surface roughness Ra, which are shown in Figure~\ref{fig:morph_params}. We found that the ragged macropores show an increased median diameter of 1.1326 $\upmu$m as compared to the unperturbed macropores with a median diameter of 0.9241 $\upmu$m. The median mesopore diameter is determined as 62.5 nm based on FIB-SEM images. In agreement with their appearance, the median surface roughness of the ragged macropores is substantially higher than that of the round macropores. The macroporosity of the sample as determined from the FIB-SEM image was 47.82 \%. The mesoporosity of the walls is with a value of 41.34\% also high, which means that roughly half of the porewalls is removed during etching. The latter value may be underestimated due to the 2D assessment. This holds for the sample selected here. For samples with longer etching times porosities of up to 70 \% were determined based on density measurements. Finally, TEM-imaging reveals that the mesopore diameter is determined as 60.06 $\pm$ 26.41 nm, relating well to the FIB-SEM characterization, and the wall thickness was calculated as 61.06 $\pm$ 30.69 nm. The specific surface area of the hp-Si presented here is determined as 502.9 $\mathrm{m}^{2}/\mathrm{g}$ which results in a surface area of 93.75 $\mathrm{m}^2$ for a 1x1x0.2 $\mathrm{cm}^3$ membrane. Additionally, the geometric tortuosity $\mathrm{\tau =}$ $\mathrm{l_{eff}/l_{0}}$, which is the ratio between the actual length of the pore, the effective transport lentgh $\mathrm{l_{eff}}$, and the virtual perfectly straight connection pathway $\mathrm{l_{0}}$ was quantified as $1.18^{+0.64}_{-0.18}$.

	\subsection{Transmission Electron Microscopy and Electron Diffraction}
\begin{figure}[H]
		\centering
		\includegraphics[width=0.45\textwidth]{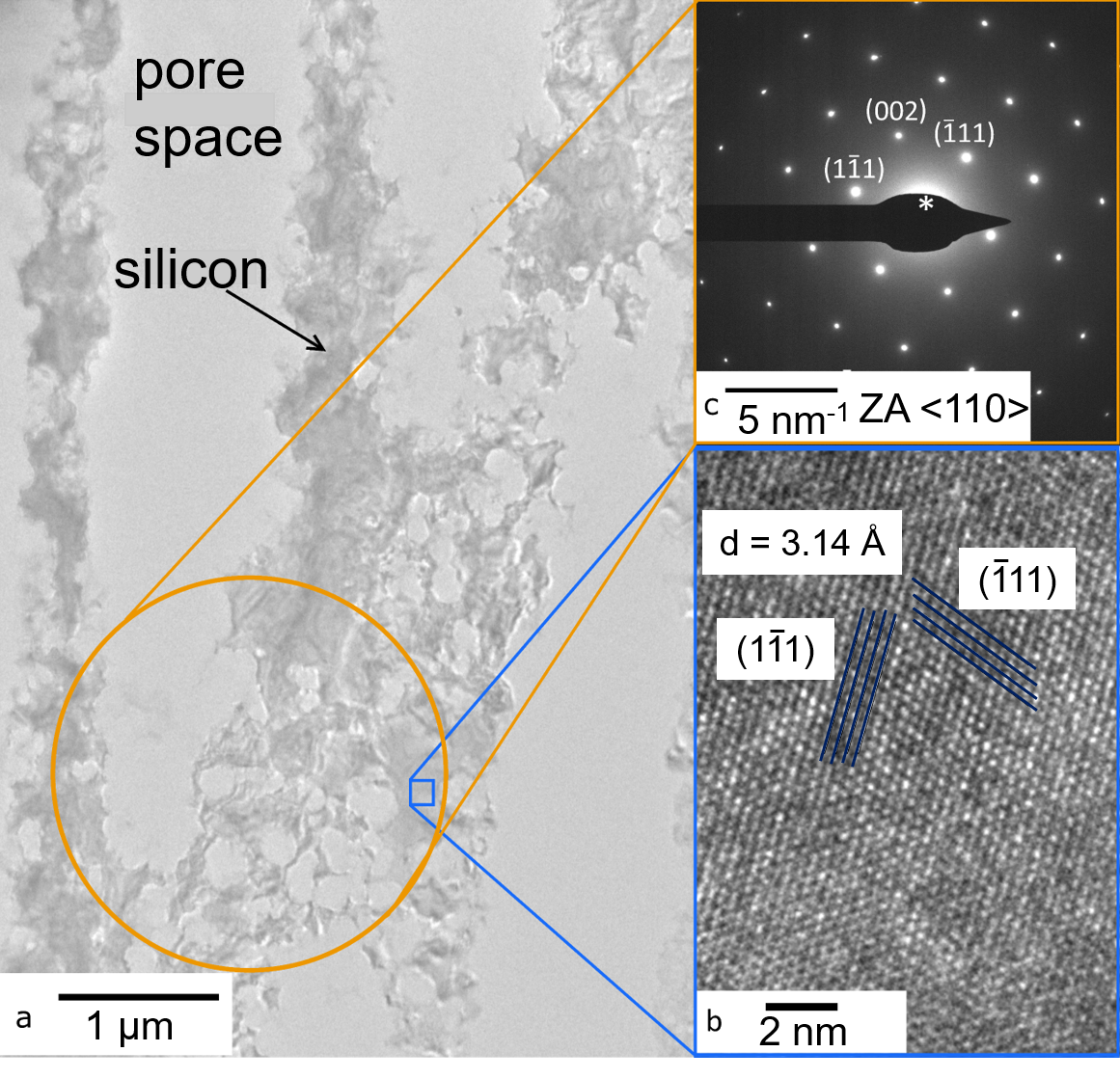}
		\caption{High-resolution transmission electron microscopy and diffraction indicates a single crystalline hp-Si structure.(a) TEM-lamella from hp-Si cutted along the macropore axis with porosified porewalls after applying MACE, (b) Magnification resolving the Si lattice with atomic resolution, (c) Electron diffraction pattern from the hp-Si structure in the zone axis <110> of Si with sharp Bragg reflexes.}
		\label{fig:TEM_TK}
	\end{figure}

	For the application of hp-Si not only the resulting porous structure but also the crystal structure after porosification plays a key-role. The starting material is a single-crystalline wafer and it is expected that it remains single-crystal during the sample preparation.
	A TEM-image of a FIB lamella, cut parallel to the macropore axis, is shown in Fig.~\ref{fig:TEM_TK}a. Electron diffraction of the porous walls result in a diffraction pattern of silicon in the zone axis <110> without diffraction rings resulting from nanocrystallinity (see Fig.~\ref{fig:TEM_TK}). High-resolution TEM imaging (\ref{fig:TEM_TK}b) shows the undisturbed, defect-free atom arrangement of silicon. 
	These results prove that hp-Si is still monocrystalline and the crystal lattice is not distorted by the metal-assisted etching. Moreover, no silver is incorporated into the lattice nor the remaining thin silicon porewalls are completely oxidized by the hydrogen peroxide. 
	
	%%%%%%%%%%%%%%%% CONCLUSION %%%%%%%%%%%%%%%%%%%%%%%%%%
	\section{Conclusion}
	We presented a novel approach based on metal-assisted chemical etching for the fabrication of hierarchical porosity in the mainstream materials silicon and silica glass. The production schemes can be adjusted in a very versatile manner in terms of lithographic pre-structuring, nanoparticle size, electrolyte constitution, etching and thermal parameters to tailor the resulting porosity. 
	
	From a more general point of view the presented fabrication scheme bridges the gap between self-organized porosity in silicon at the nanoscale with lithographically tailored porosity at the macroscale for the fabrication of a material with a large diversity of potential applications in energy storage, energy harvesting and conversion or for on-chip sensorics and actuorics.
	
	The pseudomorphic transformation and comparably small expansion upon full oxidation of the Si scaffold structure encourages us to believe that our material could also be robust against full lithiation and thus particularly interesting for lithium-ion battery applications with hp-Si as electrode material \cite{Dai2019}.
	
	For the future it could be particularly interesting to study the propelling mechanisms of the nanoparticles in the solid scaffold structure in-operando and how it can be controlled by particle size, shape and concentration or electrolyte constitution. Previous studies from \cite{Huang2011,Peng2008,Miao2017,Moumni2017} evidence, that by adjusting the electrolyte constitution and AgNP concentration the etching of different macro-structures e.g. bigger pores, larger pitches and other pore shapes should be possible and this method could be adjusted to more geometries. 
	Here, one may profit from established knowledge in the physics of active matter, specifically self-propelled Brownian particles. However, the boundary conditions, most prominently the necessary contact with a solid, the silicon dissolution process and the movement in a nanoconfined geometry are rather special and even more complex than for conventional active colloids in bulk fluids \cite{Bechinger2016,Shaebani2020,Gompper2020}. Additionally, experiments with other silicon-based materials like alloys of aluminum and silicon could be used to achieve higher complexities in the hierarchical geometries.
	
	A deeper mechanistic understanding of this physics may then allow one to tailor the self-organized porosification and thus to design the lower porosity scale of this multiscale porous material on demand. Our study also shows that a combination of self-propelled particles with porous media is not only interesting to achieve a mechanistic understanding of collective behavior of active matter, but can also be employed for tailored functional material synthesis. Thus, it evidences applications of active matter way beyond processes such as cleaning or transport of chemicals based on self-propelled nanorobots.
	
	%%%%%%%%%%%%%%%%%% METHODS AND MATERIALS %%%%%%%%%%%%%%%%%%%%%
	\section{Materials and Methods}
	
	\subsection{Materials and Chemicals}
	Monocrystalline macroporous silicon membranes doped with phosphorus (n-type) are bought from the company SmartMembranes in Halle (Saale) and serve as the starting material. The membranes have a <100> crystal orientation parallel to the surface, a pore diameter of 1  $\mu$m and a pore pitch, i.e. distance from pore centre to pore centre, of 1.5 $\mu$m. The macropores in the membrane are arranged hexagonally. The membranes are 5x5, 10x10 and 15x15 mm$^{2}$ in size and 0.2-0.3 mm in thickness. 
	The chemicals used are aqueous hydrofluoric acid (48 \%), aqueous hydrogen peroxide (30 \%) and ethanol (99.99 \% absolute for analysis) all from the supplier MERCK. The used silver nitrate powder is from Sigmar-Aldrich and has a purity of 99.98 \%.
	
	\subsection{Methods}
	The quality of the synthesized hp-Si and hp-SiO$_{2}$ is first checked with a SEM. Therefore the membranes are cut and placed in the SEM. With energy-dispersive X-ray spectroscopy during the scans the presence of silver in the membrane is analyzed. 
	To access all length scales of the hierarchical structure, transmission X-ray microscopy (TXM), transmission electron microscopy tomography (TEM-tomo) and focused ion beam combined with scanning electron microscopy (FIB-SEM) in a slice-and-view mode are used. Those tomography methods are just applied for one representative hp-Si structure.
	The TEM-tomo is performed with a needle cut from a hp-Si macropore wall which contains mesopores, whereas the FIB-SEM allows the view of a few macropores which are interconnected by some porosified walls. With the slice-and-view mode the FIB diggs into the surface of a piece of a hp-Si membrane and takes simultaneously pictures with increasing digging-depth. By reconstruction of those single 2D-images a full 3D reconstruction can be obtained.
	The TXM gives the opportunity to investigate a larger area of the hp structure and the resulting porosities.
	The combination of those tomography-methods gives a multiscale overview of the fabricated hierarchical structure. 
	
	\subsubsection{Scanning Electron Microscopy}
	SEM images of sample surfaces and fracture faces were taken with a Leo Gemini 1530 FEG-REM (Zeiss Germany). The uncoated samples were imaged using a beam voltage of 5 to 10 kV and the InLens-SE-Detector.  
	
	\subsubsection{Focused Ion Beam Tomography}
	SEM images of specimen cross-sections were taken with a Dualbeam FIB Helios NanoLab G3 UC (Thermo Fisher Scientific, USA). The sample material was vacuum infiltrated with epoxy to prevent redeposition during cross-sectioning and to support the structure during TEM sample preparation. The TLD-SE-Detector was used for imaging with a beam voltage of 2 kV. Subsequent FIB milling for slice and view tomography was done with a beam current of 25 pA for 800 10 nm thin slices. The image stack was registered and segmented after non local means filtering using Avizo (Thermo Fisher USA). The final voxel size of the reconstructed volume is 3.4x3.4x10 nm$^3$.
	
	\subsubsection{Transmission Electron Microscopy and Electron Diffraction}
	High-resolution images and electron diffraction pattern were recorded with a Talos F200X (Thermo Fisher Scientific, USA) operating at 200 kV in bright field mode. The TEM tomography experiment was performed in high-angle annular dark-field (HAADF) scanning mode also at 200 kV. The TEM lamella for imaging and electron diffraction as well as the TEM needle for TEM tomography were prepared with FIB using a standard lift-out technique.
	
	\subsubsection{Transmission X-ray Microscopy}
	The Transmission X-ray microscopy (TXM) scans were performed at the imaging beamline P05 of the PETRA III synchrotron at DESY (Hamburg) operated by the Helmholtz-Zentrum Hereon \cite{Flenner2020}, \cite{Longo2020}. A beamshaping condenser (Koehler like illumination) with 100 $\mu$m$^2$ fields, 50 nm smallest feature size and a diameter of 1.8 mm was used to illuminate the sample plane. The objective Fresnel zone plate (FZP) had an outermost zone width of 60 nm and a diameter of 250 $\mu$m. In the back focal plane of the FZP a Zernike phase ring of 1.1 $\mu$m was installed, resulting in a phase shift of $\pi$/2 at an energy of 11 keV. All optics were designed and fabricated at the Paul Scherrer Institut. A sample detector distance of 20.43 m was realized resulting in an effective pixel size of 42.9 nm using an X-ray sCMOS detector (Hamamatsu C12849-U101U).  Each tomographic scan contained 1590 projections of 0.5 s exposure time. The 3D reconstructions were performed with Tomopy, using the \textit{Gridrec} algorithm \cite{Dowd1999} in combination with a Shepp-Logan filter \cite{Gursoy2014}. After applying a binning of 2, the reconstruction lead to an isotropic voxel size of 85.8 nm. 
	
	\subsubsection{Quantitative Multiscale Porosity Analysis}
	\paragraph{TXM}
	The TXM image stack was cropped into a 400x400x400 voxel cube for further analysis. A Fourier bandpass filter with a lower and upper pixel threshold of 3 and 10 pixels was applied in Fiji/ImageJ \cite{Schindelin2009} to reduce scattering artefacts visible in the pore spaces. Subsequently, the trainable WEKA plugin \cite{Arganda-Carreras2017} in Fiji was used for segmentation of the pore space. The segmented image stack was further analysed in Avizo 2020.2 (FEI SAS, Thermo Fisher Scientific, France), where the connected pores were separated in 2D using watershedding and their barycentres were automatically calculated. The mean pore distance was determined based on the barycentres for one cross-section in Jupyter Notebook (Python 3). To this end, the mean distance of the six nearest neighbours of each pore (excluding the pore itself) were determined, as the lattice presented a triangular shape.
	
	\paragraph{FIB-SEM}
	The FIB-SEM image stack was filtered using an iterative non-local means filter \cite{Bruns2017} and subsequently cropped to a size of 10 µm x 10 µm x 300 nm. Due to strong greyscale variations along the stack, only one image from the stack was used to quantify the pore sizes within. The image was segmented semi-automatically in Avizo making use of automated region growing and manual corrections. Connected pores were separated afterwards using watershedding. Using the connected component analysis, each pore was labeled individually and its area, surface length, barycentre, diameter and image boundary connection were determined in Avizo. The resulting information was exported as .csv and imported into Jupyter Notebook for further analysis. At this stage, all pores connected to the image border were removed from the analysis to avoid any bias e.g. in terms of pore size. Using k-means (k=3) clustering the pores were classified based on their area and surface length into 1. mesopores, 2. ragged macropores, 3. round macropores. The median diameter of the three pore classes was termined and the distribution displayed as boxplots. Additionally, the surface roughness Ra for both macropore classes was determined based on the barycentres and pore edge positions. The pore edge positions were determined in Jupyter based on an image containing pore edges computed in Avizo. Finally, based on the pore classifications and volumes, the macroporosity and mesoporosity of the material were determined.
	
	\paragraph{TEM Tomography}
	The tilt series for TEM tomography consists of 161 HAADF-STEM images with a tilt spacing of 1°. The pixel size of the individual images was 384 pm. Image registration and volume reconstruction using SIRT was done with the software package Inspect3D (Thermo Fisher Scientific, USA). The volume reconstruction was segmented by thresholding in Avizo. The convex hull of the pore space was then determined automatically in Avizo. Using the BoneJ plugin \cite{Doube2010} in Fiji, the pore size and wall thickness of the resulting geometry were determined, respectively. Additionally, the pore space was skeletonized in Fiji and analyszed using BoneJ, to determine its geometric tortuosity \cite{Clennell1997}. Geodesic tortuosity measures the lengths of shortest transportation paths with respect to the materials thickness \cite{Neumann2019Tortuosity}. It was calculated for all branches of the skeleton longer than 10 voxels (3.83 nm).
	
	%%%%%%%%%%%%%%%%%%%% brACKNOWLEDGE %%%%%%%%%%%%%%%%%%%%%%%%%%%%%%
	
	\section{Acknowledgements}
	We thank Deutsches Elektronen-Synchrotron DESY Hamburg for access to the beamline P05 of the PETRA III synchrotron. We thank Dr. Elena Longo for performing the transmission X-ray microscopy experiments TXM at P05. S.G. was supported by the Deutsche Forschungsgemeinschaft (DFG) within the Collaborative Research Centre CRC 986 “Tailor-Made Multi-Scale Materials Systems” project number 192346071. P.H. gratefully acknowledges support by the DFG Graduate School GRK 2462 ”Processes in natural and technical Particle-Fluid-Systems (PintPFS)” (Project No. 390794421). This project has also received funding from the European Innovation Council (EIC) under the European Union's Horizon 2020 research and innovation program under grant agreement No. 964524 EHAWEDRY: "Energy harvesting via wetting-drying cycles with nanoporous electrodes" (H2020-FETOPEN-1-2021-2025). We also acknowledge the scientific exchange and support of the Centre for Molecular Water Science CMWS, Hamburg (Germany).\\ 
	Author contributions: S.G., M.B. and P.H. conceived the project. S.G. and M.B. performed the synthesis. S.G. performed the SEM measurements. The FIB-SEM and TEM were taken by T.K. and D.R. and the TXM was performed by I.G. and E.L. B.Z.-P. evaluated the tomography data-sets from TXM, FIB-SEM and TEM-tomography. S.G., M.B. and P.H. wrote the manuscript. All other authors proofread the manuscript. Competing interests: The authors declare that they have no competing interests. Data and materials availability: All data needed to evaluate the conclusions in the paper are present in the paper and/or the Supplementary Materials.
	We also acknowledge that the synthesis pathway presented here is a pending patent with the number DE 10 2020 124 532.7.
	
	%%%%%%%%%%%%%%%%%% REFERENCES %%%%%%%%%%%%%%%%%%%%%%%%%%%%%%%% 
	%\addbibresource{MACE.bib}
	%\addbibresource{HuberLabMendeleyBib.bib}

\end{document}